\documentstyle[12pt,epsf,epsfig]{article}
\begin{document}

\pagestyle{empty}
\baselineskip=21pt
\rightline{ACT-02/11}
\rightline{CERN-TH/2002-370}
\rightline{LAPTH-957/02}
\rightline{MIFP-02-10}
\rightline{\tt hep-th/0212272}

\begin{center}
 {\large \bf Universal Calabi-Yau Algebra: }\\
{\large \bf Classification and Enumeration of Fibrations}\\
\end{center}

\vspace{.03in}

\begin{center}
{\bf F. Anselmo}$^1$, {\bf J. Ellis}$^1$, {\bf D. V. Nanopoulos}$^2$ and 
{\bf G. Volkov}$^{1,3}$

\vspace{.02in}

$^1${\it CERN, CH-1211 Geneva, Switzerland}  \\
$^2${\it George P. and Cynthia W. Mitchell Institute of Fundamental 
Physics, \\
Texas A \& M University, College Station, TX 77843-4242, USA \\
Astroparticle Physics Group, Houston Advanced Research Center (HARC),\\
The Mitchell Campus, The Woodlands, TX 77381, USA \\
Chair of Theoretical Physics, Academy of Athens, Division of Natural 
Sciences, \\
28 Panepistimiou Avenue, Athens 10679, Greece}\\
$^3${\it LAPP TH, Annecy-Le-Vieux, France, and\\
  St Petersburg Nuclear Physics Institute, Gatchina, 188300 St Petersburg,
Russia\\}


\vspace{.25in}

{\bf Abstract}
\end{center}

We apply a universal normal Calabi-Yau algebra to the construction and
classification of compact complex $n$-dimensional spaces with $SU(n)$
holonomy and their fibrations. This algebraic approach includes natural
extensions of reflexive weight vectors to higher dimensions and a `dual'
construction based on the Diophantine decomposition of invariant
monomials. The latter provides recurrence formulae for the
numbers of fibrations of Calabi-Yau spaces in arbitrary 
dimensions, which we exhibit explicitly for some Weierstrass and K3 
examples.

\vspace{0.25in}
\leftline{CERN-TH/2002-370}
\leftline{December 2002}

\newpage

\section{The Algebraic Way to Unify Calabi-Yau Geometry}

Geometrical ideas play ever-increasing r\^oles in the quest to unify all
the fundamental interactions. They were introduced by Einstein in the
formulation of general relativity, and extended to higher dimensions by
Kaluza and Klein in order to include electromagnetism. To explain the
appearance of electromagnetism, it was enough to introduce
just one extra dimension with the topology of a circle, exploiting the
geometrical equivalence between the $U(1)$ gauge group and the circle. In
order to find a corresponding geometrical origin for the full Yang-Mills
symmetries of the Standard Model, namely $SU(3) \times SU(2) \times U(1)$,
one needs to consider more complicated geometrical structures.

The modern approach to gauge symmetry is based on string theory, whose
underlying geometrical nature is still mysterious, but includes an
enormous extension of general coordinate invariance in ten or eleven
dimensions. This is large enough to include the gauge symmetries of the
Standard Model in four dimensions, if one compactifies six surplus
dimensions on a suitable complex three-dimensional space, called a
Calabi-Yau manifold~\cite{CY}. In the original compactifications of weakly-coupled 
ten-dimensional heterotic string theory, the resulting four-dimensional 
gauge group would be some subgroup of $E_6$.

This construction has since been extended to various non-perturbative
constructions. It has been realized that additional gauge-group factors
may appear in suitable singular limits of the Calabi-Yau manifold~\cite{Kodaira}.  
For example, one approach to
non-perturbative string theory is based on twelve-dimensional $F$ theory,
which may be compactified to four dimensions on a complex four-dimensional
space with $SU(4)$ holonomy. We also note that many $CY_3$ or $CY_4$
spaces can be obtained as Complete-Intersection Calabi-Yau (CICY) spaces,
i.e, as projections inside higher-dimensional $CY_n$, motivating further
studies of the latter also for $n > 4$.

These geometrical ideas exemplify the physics interest of classifying 
systematically spaces with holonomy groups in the series $SU(n), SO(n)$ 
and $Sp(n)$, as well as $G_2$ and Spin(7). Listings are available of 
special cases such as $K3$ spaces with $SU(2)$ holonomy and $CY_3$ spaces 
with $SU(3)$ holonomy~\cite{K3,Skarke}, and there are also many results for 
other holonomy 
groups~\cite{berger,Joyce}. Ideally, one would like to classify 
these spaces in a systematic 
way, much as Cartan provided an algebraic classification of Lie groups~\cite{Cartan}.
This is, of course, a very ambitious programme, for which only partial 
results are available.

We have proposed an algebraic approach~\cite{AENV1,AENV2} 
to the problem of classifying
complex $CY_n$ manifolds with $SU(n)$ holonomy, which is based on their
identifications with the loci of zeroes of polynomials in suitable complex
projective spaces, and their complete intersections. These complex
projective spaces in different dimensions are characterized by `reflexive'
projective weight vectors $\vec k$. Our approach has been based on the
systematic extension of lower-dimensional projective vectors to higher
dimensions, and their combination via binary, ternary, quaternary, etc.,
algebraic operations of increasing `arity'.

We have recently proposed~\cite{AENV3} 
a supplement to this approach which is based on
a `dual' approach via the monomials $x^{{\vec{\mu}}_{\alpha}} =
{x_1^{{\mu}_{1\alpha}}
x_2^{{\mu}_{2\alpha}}...x_{n+1}^{{\mu}_{(n+1)\alpha}}}$ in the
quasihomogeneous coordinates ${x_1,...,x_{n+1}}$ that obey a `duality'
condition ${\vec {\mu}}_{\alpha} \cdot \vec {k}\,=\,[d]$. 
Specifically, we
showed how $CY_n$ spaces could be obtained by the Diophantine
decomposition of simple invariant monomials, a technique that gives
immediate insights into the fibrations of higher-dimensional Calabi-Yau
spaces involving lower-dimensional Calabi-Yau spaces.

In this Letter, we summarize briefly the essential aspects of this new
Diophantine algebraic approach to the systematic classification of
Calabi-Yau spaces, demonstrating its complementarity to our previous
expansion technique. In particular, we present a number of explicit
results for the numbers of fibrations of Calabi-Yau spaces in arbitrary
numbers of complex dimensions. These results support our claim to have
formulated a `Universal Calabi-Yau Algebra'~\cite{Burris}
capable in principle of
decoding the full Calabi-Yau genome. Moreover, as we comment at the end,
the techniques used here could in principle also be used to classify the
series of spaces with $SO(n)$.

\section{The Arity-Dimension Structure of Universal
Calabi-Yau Algebra}

The starting point for our algebraic classification of Calabi-Yau 
spaces
has been the construction of `reflexive' weight vectors ${\vec k}$, 
whose
components specify complex quasihomogeneous projective spaces
${CP^n(k_1,k_2,...,k_{n+1})}$. These have {$(n+1)$} quasihomogeneous
coordinates ${x_1,...,x_{n+1}}$, which are subject to the following
identification:
\begin{equation}
(x_1, \ldots ,x_{n+1})\,\sim\, (\lambda ^{k_1} \cdot x_1, \ldots ,
\lambda ^{k_{n+1}} \cdot x_{n+1}).
\label{quasihom}
\end{equation}
A general quasihomogeneous polynomial of degree $[d]$ is a linear
combination
\begin{equation}
{\wp} = \sum_{\vec{\mu}_{\alpha}} c_{\vec{\mu}_{\alpha}}
x^{\vec{\mu}_{\alpha}}
\label{poly2} 
\end{equation}
of  monomials  $x^{{\vec{\mu}}_{\alpha}} = {x_1^{{\mu}_{1\alpha}}
x_2^{{\mu}_{2\alpha}}...x_{n+1}^{{\mu}_{(n+1)\alpha}}}$
with the condition:
\begin{equation}
{\vec {\mu}}_{\alpha} \cdot \vec {k}\,=\,[d].
\label{poly3}   
\end{equation}
A $d$-dimensional Calabi-Yau space $X_d$ can be given by the locus of
zeroes of a transversal quasihomogeneous polynomial ${\wp}$ of degree 
$deg
({\wp})=[d]: [d] = \sum _{j=1}^{n+1} k_j$ in such a complex projective
space $CP^n(\vec {k}) \equiv CP^n (k_1,...,
k_{n+1})$~\cite{K3}:
\begin{eqnarray}
X \equiv X^{(n-1)}({k}) \equiv
\{\vec{x}=(x_1,...,x_{n+1})\in CP^n({k})|{\wp}(\vec{x}) = 0\}.   
\label{poly1}
\end{eqnarray}
This algebraic projective variety is irreducible if and only if  
its polynomial $\wp$ is irreducible. A hypersurface will be smooth
for almost all choices of polynomials. To obtain Calabi-Yau $d$-folds, one
should choose reflexive weight vectors (RWVs) related to
reflexive Batyrev polyhedra~\cite{Batyrev}.
Other examples of compact Calabi-Yau manifolds can be obtained as the
complete intersections (CICY) of such quasihomogeneous polynomial
constraints:
\begin{eqnarray}
X^{(n-r)}_{CICY}
=\{\vec{x}=(x_1,\ldots x_{n+1})\in CP^n \,|\, {\wp}_1(\vec x)
=\ldots ={\wp}_r(\vec x)=0 \},
\end{eqnarray}
where each polynomial ${\wp}_i$ is determined by some extended weight vector
$\vec{k}_i$, $i=1,\ldots, r$, where $r$ is the arity

\begin{figure}[th!]
   \begin{center}
   \mbox{
   \epsfig{figure=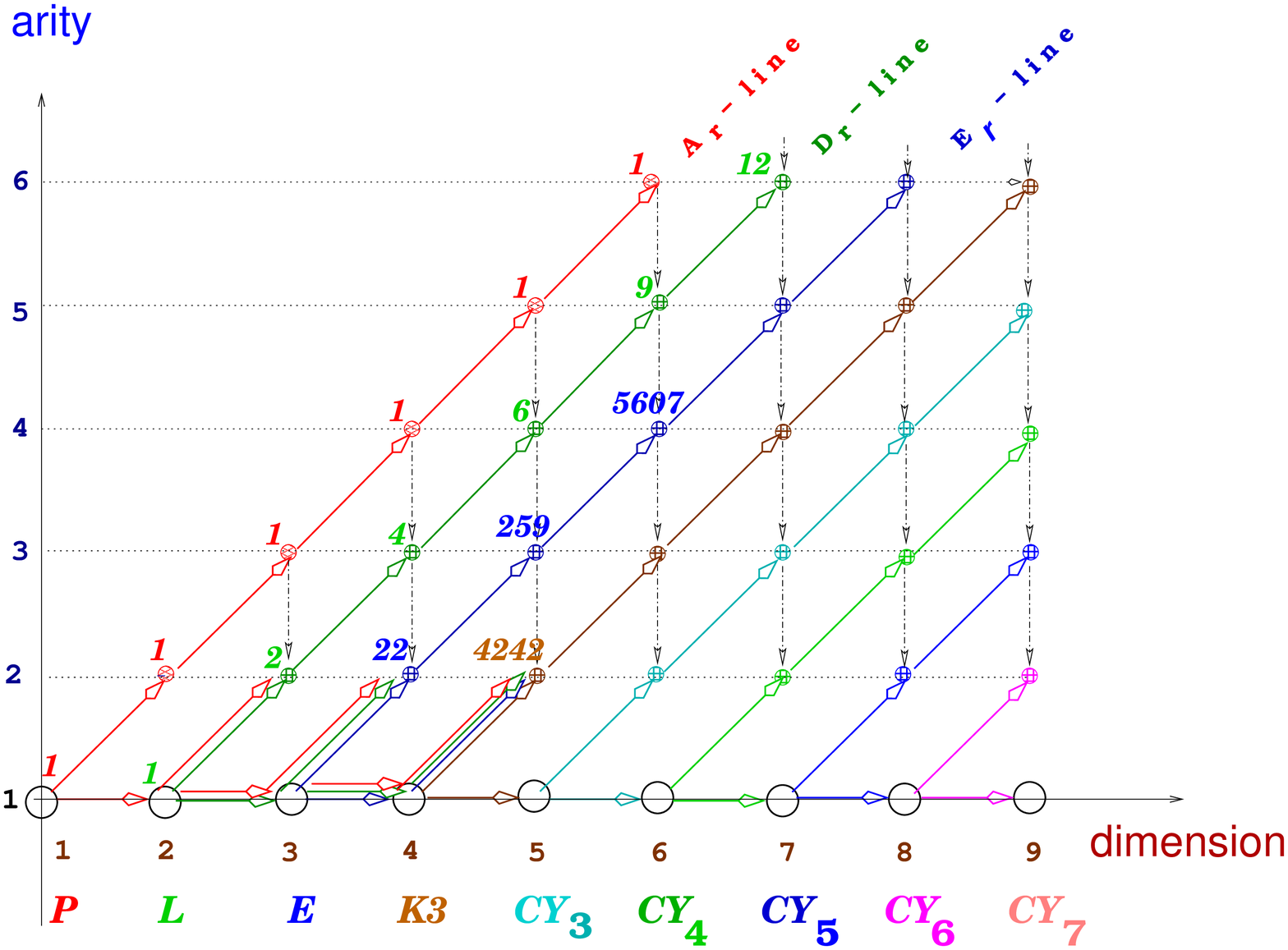,height=8cm,width=9cm}}
   \end{center}
   \caption{\it
The arity-dimension plane for complex manifolds with $SU(n)$ holonomy, 
showing the numbers of eldest
vectors/chains obtained by normal extensions of RWVs, including complete
results for $CY_3$ and lower-dimensional spaces, and partial results for
$CY_4$ and $CY_5$ spaces.}
\label{basmon1}
\end{figure}

These RWVs that specify the polynomials $\wp_i$ may be classified using 
 the natural extensions of lower-dimensional vectors and their
combinations via binary, ternary, etc., operations $\omega_r$, as
illustrated in Fig.~\ref{basmon1}.  The Universal Calabi-Yau Algebra
(UCYA) structure of reflexive weight vectors (RWVs) in different
dimensions depends on two integer parameters: the dimension $n$ of the
RWVs, and the {\it arity} $r$ of the combination operation $\omega_r$.

A useful technique for constructing Calabi-Yau spaces in any number of
dimensions is to visualize the various possible monomials
$(x_1^{\mu_1}x_2^{\mu_2}...x_n^{\mu_{n+1}})_{\alpha} $ as points
${m}_{\alpha}=(\mu_1,...,\mu_{n+1})_{\alpha}$ in the $Z_{n+1}$ integer
lattice of an $n$-dimensional polyhedron~\footnote{Using this technique,
Batyrev~\cite{Batyrev} demonstrated by explicit construction how to
associate a mirror polyhedron to each Calabi-Yau space. This approach also
established in a very elegant way the corresponding mirror duality among
Calabi-Yau spaces, in which CICY spaces play an essential role.}.
This feature enabled us to introduce a complementary algebraic approach to
the construction of Calabi-Yau spaces, based on the construction of
suitable monomials ${\vec \mu}$ obeying the `duality' condition: $\vec{k}
\cdot \vec{\mu}_{\alpha} = d$. This construction supplements the previous
geometrical method related to Batyrev polyhedra, and enables one to
calculate the numbers of eldest vectors, and hence chains, in arbitrary
dimensions. We have verified explicitly that the eldest vectors found in
the two different ways agree in several instances for both $CY_3$ and
$CY_4$ spaces~\cite{AENV3}, providing increased confidence in our results. 
The study of the Calabi-Yau equations and the associated hypersurfaces via the
remarkable composite properties of invariant monomials (IMs) provides an 
algebraic alternative to reflexive polyhedron techniques.

Central r\^oles are played in this approach by composite lower-dimensional
structures within $CY$ $d$-folds, which can be seen by the algebraically 
dual ways of
expansions using weight vectors $\vec k$ and IMs. By
analogy with the Galois normal extension of fields, we term the first way
of expanding weight vectors a {\it normal} extension, and the dual
decomposition in terms of IMs we call the {\it Diophantine} expansion.
These two expansion techniques are consistently combined in our algebraic
approach, whose composition rules exhibit explicitly the internal
structure of the Calabi-Yau algebra. Our method is closely connected to
the well-known Cartan method for constructing Lie algebras, and reveals
various structural relationships between the sets of Calabi-Yau spaces of
different dimensions. We interpret our approach as revealing a `Universal
Calabi-Yau Algebra'~\cite{Burris} for the following reasons: `Universal'
because it may, in principle, be used to generate all Calabi-Yau manifolds
of any dimension with all possible substructures, and `Algebra' because it
is based on a sequence of binary and higher $n$-ary operations on weight
vectors and monomials.
 
This Universal Calabi-Yau Algebra (UCYA) acts upon the set of reflexive
weight vectors in all dimensions, $A_n \equiv \{$RWV$(n) \}$, and the
corresponding set of invariant monomials, $\{IM(n) \}$, which is `dual'
to $A_n$ in the sense of (\ref{poly3}). 
The IMs are the minimal set of monomials determining the eldest vector 
and its chain, and thence the full list of
weight vectors in the corresponding chain. 
According to their degrees $a = 2, 3, 4, ...$
we term them conics, cubics, etc.. The structure of each sloping line in 
the dimension-arity plane is determined by the corresponding
set of IMs, e.g., the first $A_r$ line is determined by the unit monomials 
$E_n$, the second by conics and linear monomials, the third line by
cubics and quartics, taking into account also conics, etc..
The number of IMs is
much less the full set of monomials ${\vec m}_{\alpha}: 1 \leq \alpha \leq
\alpha_{max}$ that determine the Calabi-Yau equation.
To construct them, we can start from the
unit IM in some dimension $n$ and then, via a Diophantine expansion,
generate related conic IMs, cubic IMs, quartic IMs, etc..
This process may then be continued by studying in turn the
Diophantine expansions of conic IMs, of cubic IMs, etc..

We note in addition that the algebraic-geometry realization~\cite{K3,Can1}
of Coxeter-Dynkin diagrams provides a general characterization of the
possible structures in singular limits of Calabi-Yau hypersurfaces, which
are associated with possible gauge groups.  Thus, a deeper understanding
of the origins of gauge invariance provides an additional motivation for
studying string vacua via our unification of the complex geometry of $d=1$
elliptic curves, complex tori, $K3$ manifolds, $CY_3$, $CY_4$, etc. This
point is illustrated in Figs.~\ref{basmon1}, where the points on the the
first three sloping lines, labelled $A_r$ (red), $D_r$ (green) and $E$
(blue), correspond to those $d$-folds that are characterized by the
`maximal' quotient $A, D, E$ singularities, respectively~\footnote{To be
more precise, the $D$ line includes also $A$-type singularities, and the
$E$ line includes also $D$-type and $A$-type singularities.}. This
characterization of the types of singularities is directly connected to
the degrees of the associated monomials - linear, conics, cubics,
quartics, etc., that appear along the corresponding sloping lines.

In summary: the UCYA provides a two-parameter classification of
$CY_d$ spaces in terms of arity $r$ and dimension $n$, which is based on
the following ingredients:

\begin{itemize}
\item{Universal composition rules,}
\item{Normal expansions and Diophantine decompositions,}
\item{Mirror symmetry,}
\item{Singularities and their links with Cartan-Lie algebras.}
\end{itemize}

We have shown that this algebraic approach leads us to a natural formalism
for a unified description of complex geometry in all dimensions, including
$K3$ spaces and Calabi-Yau $d$-folds for any $d$~\cite{AENV1,AENV3}.

As an example of the extension procedure for RWVs, as applied to $K3$
manifolds, we classified~\cite{AENV1} the 95 different possible weight
vectors $\vec{k}$ in 22 binary chains generated by pairs of extended
vectors, which included 90 of the total, and 4 ternary chains generated by
triplets of extended vectors, which yielded 91 weight vectors, of which 4
were not included in the binary chains. The one remaining $K3$ weight
vector was found in a quaternary chain~\cite{AENV1}. This algebraic
construction provides a convenient way of generating all the $K3$ weight
vectors, and arranging them in chains of related vectors whose overlaps
yield further indirect relationships.

\section{The Classification and Enumeration of Fibrations}

\begin{figure}[th!]
   \begin{center}
   \mbox{
   \epsfig{figure=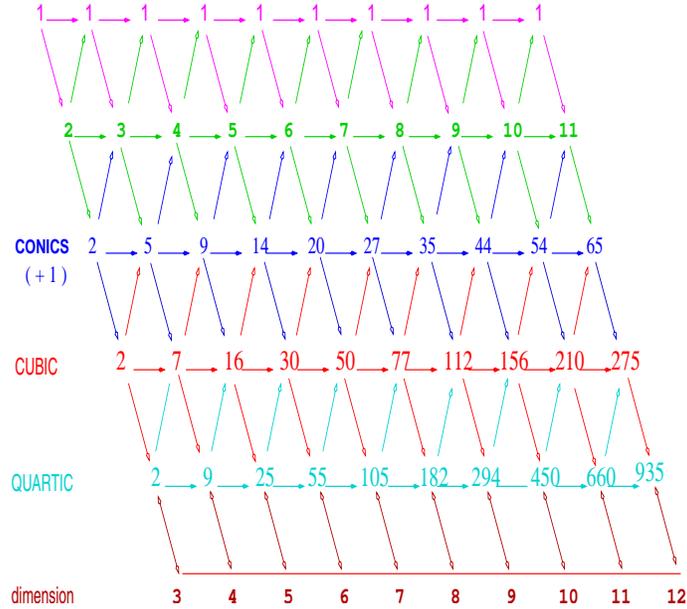,height=8cm,width=9cm}}
   \end{center}
   \caption{\it
Lattice illustrating recurrence relations for the numbers
of conic, cubic and quartic monomials.}
\label{recur}
\end{figure}

We now show how our technique for building higher-dimensional Calabi-Yau 
spaces
systematically out of lower-dimensional ones enables us also to enumerate
explicitly their fibrations. As examples, we showed
previously~\cite{AENV1,AENV2,AENV3} how our construction reveals elliptic 
and
$K3$ fibrations of $CY_3$ manifold~\footnote{Our approach may also be used
to obtain the projective weight vector structure of a mirror manifold,
starting from those of a given Calabi-Yau manifold.}. We now present some
further results derived via the new description of $CY_d$ spaces based on
the structures of the set of invariant monomials (IMs). Recurrence
relations for conic, cubic and quartic monomials give us the formulae for
the numbers of IMs and hence fibrations in arbitrary dimensions, leading
us to a complete solution for the fibrations of $CY_d$ spaces along the
$D_r$- and $E_r$-lines in Fig.~\ref{basmon1}. These results confirm that,
in the framework of the UCYA, the Calabi-Yau `genome' can in principle be
solved completely. As we explain in more detail below, the IMs 
determine completely the fibration structures 
of the 22 $K3$ chains mentioned earlier, as shown in Fig.~\ref{recur}:

\begin{eqnarray}
\{IM\}_4 &\mapsto&\biggl ( 1\cdot\{4\}_{\Delta} \biggl)+
\biggl({\bf 2\cdot\{10\}_{\Delta}}\biggl ) \nonumber\\
&+&\biggl (2 \cdot \{5\}_{\Delta}+ 1 \cdot \{5\}_{\Box} \biggl )\nonumber\\
&+&\biggl ({\bf4 \cdot\{9\}_{\Delta}}+ 2 \cdot \{9\}_{\Box} \biggl ) \nonumber\\
&+&\biggl({\bf 7\cdot\{ 7\}_{\Delta}} +1\cdot\{ 7\}_{\Box}\biggl )\nonumber\\
&+& \biggl(1 \cdot \{ 6\}_{\Box}\biggl )
+\biggl(1 \cdot \{ 8\}_{\Box}  \biggl )\nonumber\\
&\mapsto & \{22\},
\end{eqnarray}
where we label planar sections through Batyrev polyhedra via the number of 
points they contain and their geometric shapes: $\{ N\}_{\Delta, \Box}$, 
etc.. This expansion in terms of fibration structures for
$K3$ spaces may be extended to more general $CY_d$ spaces, via recurrence
relations. Each of the terms $\{10,4,...\}_{\Delta,
\Box, ...}$ in the expansion has its own recurrence relation, of which we
give below several examples, namely those indicated in bold script above:
${\bf 2\cdot\{10\}_{\Delta}}$, etc., providing complete results in
any number of dimensions for the numbers of $CY_d$ spaces with these particular
fibrations. A similar recurrence formula could in principle be derived for 
any analogous fibration.

In the five-dimensional case corresponding to $CY_3$ spaces, we have 
derived the types and numbers of IMs which determine the structures of
the full 259 (161 irreducible) chains, which are similar to those for
the $K3$ case above:
\begin{eqnarray}
\{IM\}_5&\mapsto&\biggl ( 9\cdot\{4\}_{\Delta} +
{\bf 4\cdot\{10\}_{\Delta}}\biggl )
 \nonumber\\
&+&\biggl (16 \cdot \{5\}_{\Delta}+ 5 \cdot \{5\}_{\Box}+
1 \cdot \{5\}_{\Box'} \biggl )\nonumber\\
&+&\biggl ({\bf 11\cdot\{9\}_{\Delta}}+ 5 \cdot \{9\}_{\Box}+
1 \cdot \{9\}_{\Box'}\biggl ) \nonumber\\
&+&\biggl({\bf 28\cdot\{ 7\}_{\Delta}} +7\cdot\{ 7\}_{\Box}+
1\cdot\{ 7\}_{Quint} \biggl )\nonumber\\
&+& \biggl(8 \cdot \{ 6\}_{\Box} +1 \cdot \{ 6\}_{Quint}\biggl )\nonumber\\
&+&\biggl(6 \cdot \{ 8\}_{\Box} +1 \cdot \{ 8\}_{Quint} \biggl )
\nonumber\\
&\mapsto& \{161\}
\end{eqnarray}
In the six-dimensional case corresponding to $CY_4$ spaces, out of the
5,607 6-dimensional 4-vector chains, just 2,111 are independent. We find
the following classification of their fibrations:

\begin{eqnarray}
\{IM\}_6&\mapsto&\biggl ( 37\cdot\{4\}_{\Delta} +
{\bf7\cdot\{10\}_{\Delta}}\biggl ) \nonumber\\
&+&\biggl (66 \cdot \{5\}_{\Delta}+ 27 \cdot \{5\}_{\Box}+
6 \cdot \{5\}_{\Box'} \biggl )\nonumber\\
&+&\biggl ({\bf24\cdot\{9\}_{\Delta}}+ 11 \cdot \{9\}_{\Box}+
5 \cdot \{9\}_{\Box'}\biggl ) \nonumber\\
&+&\biggl({\bf 84\cdot\{ 7\}_{\Delta}} +28\cdot\{ 7\}_{\Box}+
5\cdot\{ 7\}_{Quint}+1\cdot\{ 7\}_{Sixt} \biggl )\nonumber\\
&+& \biggl(36 \cdot \{ 6\}_{\Box} +5 \cdot \{ 6\}_{Quint}\biggl )\nonumber\\
&+&\biggl(21 \cdot \{ 8\}_{\Box} +5 \cdot \{ 8\}_{Quint} \biggl )\nonumber\\
&\mapsto& \{2111\}.
\end{eqnarray}
Similar expressions can be derived for any desired dimension.

Illustrating the derivation, we recall
that the first $A_r$ line in Fig.~\ref{basmon1} is characterized by 
the unit monomials $E_n=(1,...,)$. The $CY_d$ spaces along the 
second $D_r$ line with arity $r = (n - 1)$ have `almost trivial' 
substructures, i.e., `circles', whose reflexive polyhedra are linear.
These may be classified and enumerated by Diophantine 
expansions of the unit monomials $E_n=(1,...,1)_n$
in terms of pairs of
conics $C_{i(n)}$ and
$C_{j(n)}$, which should satisfy the following Diophantine
property:
\begin{equation}
\frac{1}{2} (C_{i(n)}+C_{j(n)})=E_n,
\label{conicsplit}
\end{equation}
where the index $n$ notes the dimension being considered.
This Diophantine expansion yields the following numbers
of possible different types of conic monomials in any dimension $n$,
\begin{equation}
N_{conics}=\frac{(n)(n-1)}{2},
\label{conics}
\end{equation}
as may easily be shown by induction. As we discuss later,
the structures of the $CY_d$ spaces on the next third $E_r$ line are 
correspondingly determined by Diophantine expansions in terms of cubic
and quartic IMs, either directly:
$(P_1+P_2+P_3)/3=E_n$, or in two steps: $(Q_{j1}+Q_{j2})/2=C_j$ where 
$\vec{k}^{ext} \cdot Q_a=d$ and $\vec{k}^{ext} C_j=d$.

In order to enumerate the IMs and the corresponding chains of Calabi-Yau
spaces, which are given by suitable pairs (\ref{conicsplit}) of conics, 
one solves the following Diophantine equations:
\begin{equation}
\vec{k}^{i,ext} \cdot C_{1(n)}=\vec{k}^{i,ext} \cdot
E_n=d(\vec{k}^{i,ext}).  
\end{equation} 
To give sense to these equations
and, consequently, to evaluate the finite numbers of chains and their 
eldest vectors in the case of arity
$r = (n - 1)$, we first recall that, in the UCYA, the points on this
line in the arity-dimension plane are determined by $n$-dimensional 
extensions of the two eldest
vectors $\vec{k}_1=(1)$ and $\vec{k}_2=(1,1)$.  This means that the
possible values of $d(\vec{k}^{i(ex)})$ in these equations are only 1 and
2, whereas the components of the extended vectors can only be 0 or 
1. Due our algebra, this second sloping line is determined also only by 
extensions of the weight vectors (1) and (1,1), so their components can 
include only one or two units.
It is then simple to verify by induction the following recurrence formula
for the numbers of chains along the second diagonal line in the 
arity-dimension plane: 
\begin{eqnarray}
N_{chains}&=&k\cdot(k+1), \qquad if \;\,\,n=(2k+1) \nonumber\\ N_{chains}&=&
k^2, \qquad if \;\,\,n=(2k). \nonumber\\ 
\end{eqnarray} 
Thus, along the line $r = (n - 1)$, the numbers of the eldest vectors and 
chains in dimensions $n = 2, 3, 4, ...$ are the following: 1, 2, 4, 6, 9, 
12, 16, 20, 25, 30, 36, 42, 49, 56, 64, 72, 81, 90, 100, 110, 121, 132, 
144, .... 

Extending our previous approach to the third line in Fig.~\ref{basmon1},
the first step is to enumerate the cubic and quartic monomials, from which
we can find all the IMs along this $E_r$ line. 
The appearance of cubic
monomials is connected with the following new Diophantine condition for
the expansion of the unit monomials $E_n$ of the $A_r$ line:
\begin{eqnarray}
E_n \mapsto \{P_1,P_2,P_3|\frac{1}{3}(P_1+P_2+P_3)=E_n\}.
\end{eqnarray}
However, the set of appropriate cubic monomials is somewhat more
restricted. Similarly, the appearance of quartic monomials is connected
with the possible Diophantine expansion of the conic monomials $C_{i(n)}$
of the second $D_r$ line:
\begin{eqnarray}
C_{i(n)}\, \mapsto\, \{P_1,P_2|\frac{1}{2}(P_1+P_2)=C_{i(n)}\}.
\end{eqnarray}
Again, there are some further restrictions on the list of possible quartic
monomials, which we do not discuss here. 

As indicated in Fig.~\ref{recur}, there are
recurrence formulae for the numbers of monomials in any dimension, which
are obvious for the leading (red and green) lines in the arity-dimension 
plane. The resulting expressions for the numbers of cubic and quartic 
monomials are, respectively:
\begin{eqnarray}
N_{cubics}&=&\frac{1}{6}(n-2) (n-1) (n+3)\nonumber\\
N_{quartics}&=&\frac{1}{24} (n-2)(n-1)(n)(n+5)\nonumber\\
\end{eqnarray}
There are remarkable links between the numbers of conics, cubics and
quartics. For example, to obtain the number of quartics in dimension $n$,
one should sum over all the cubics in dimensions $3, 4, ..., n$, i.e.,
$N_{Quart}^{(n)}={\sum_{i=3}^{i=n}} N_{Cub}^{(i)}$. Thus, as seen in
Fig.~\ref{recur}, the number 105 of quartic monomials in the septic
Calabi-Yau case can be represented as follows: $2_{dim=3}+7_{dim=4}+
16_{dim=5}+ 30_{dim=6}+50_{dim=7}$.

Based on Fig.~\ref{recur}, one can convince oneself that there also
exist $n$-dimensional recurrence formulae for the numbers of IMs 
along other diagonal lines in any dimension, as
we have found for the first two lines on the arity-dimension plot in
Fig.~\ref{basmon1}. However, the situation can become complicated,
because, in the construction of the cubic and quartic IMs, one must also
take into account conic and conic + cubic monomials, respectively.
In the case of Calabi-Yau spaces with Weierstrass fibres, it is also
important to know the list of sextic monomials, 
which is given by the
following recurrence formula:
\begin{eqnarray}
C_{n+2}^{n-3}= 
\frac{(n+2)!}{(n-3)!5!},
\end{eqnarray}
where $n \geq 3$ is the dimension of the weight-vector space. Via these 
sextic monomials and three- and four-fold Diophantine decompositions of 
the unit vector $E_n$, we then obtain the following expression for the 
number of Weierstrass IMs, $\{3\}$ and $\{4\}$:
\begin{equation}
N_{W}(n) = N_{\{7\}_\Delta}(n) = C_{n+3}^{n-3} = \frac{(n+3)!}{(6!)(n-3)!},
\end{equation}
valid for all dimensions.

Similarly, one can find a recurrence relation for
$\{IM\}=\{{9}\}_{\Delta}$, which is constructed from two quartic
monomials, one conic and $E_n$. In this case the difference of the two
quartic monomials should be divisible by four. Taking into account all
possible quartics, after some effort, one can find the following formula
for the number of these IMs:
\begin{eqnarray}
N_{\{9 \}_\Delta}(n) = \frac{1}{3} \cdot (n-2)(n^2-4n+6).
\end{eqnarray}   
This expression gives the following numbers:  $1, 4, 11, 24, 45, 76, 119,
176, 249, ...$ for $n = 3, 4, 5, 6, 7, 8,$ $9, 10, 11, ..$, respectively.
We find analogously the recurrence relation
\begin{equation}
N_{\{ 10 \}_\Delta}(n)=N_{\{ 10 \}_\Delta}(n-1)+n(3),
\end{equation}
where $n(3)$ denotes the number of the ways of decomposing $n$ into 
three positive integers.

This way of getting the recurrence formula for Calabi-Yau spaces with 
elliptic fibres
$\{10\}_{\Delta}$ can be extended to the cases of $CY_d$ spaces with $K3$
fibres, such as those described by $\vec{k}_4=(1,1,1,1)[4]$ whose 
algebraic equation includes 35 monomials.
The ${IM}_4$ for this $K3$ space contains the
four quartic monomials $P_1,P_2,P_3,P_4$ obeying the Diophantine equation:
$(P_1+P_2+P_3+P_4)/4=E_4$. These monomials have in addition one very important
condition: $P_i - P_j$ should be divisible by {4} for each choice of {$i,
j = 1, 2, 3, 4, i \neq j$}. The types of different $n$-dimensional
$IM_4$, describing the $CY_d: n=d+2 \geq 4$ spaces with such
$\{35\}_{\Delta}$ fibres are constructed only from the
numbers {4} and {0}, and possibly the unit.
Similarly to the case of the third $E_r$ sloping line,
the recurrence formulae for these {IMs} is determined by the
expansions of positive integer numbers in terms of four positive 
integers. Indeed, for each slope line there is a recurrence formula of the 
type 
\begin{equation}
N_{...\Delta}(n)=N_{...\Delta}(n-1)+n(p),
\end{equation}
where $p$ is  the number of the sloping line, and 
$N_{...\Delta}(n_{min})=2: n_{min}\geq 3$. The numbers of many 
other desired IMs can be established in a similar way.

We give finally the example of $CY_d$ spaces on the 
fourth ($K3$) line with
an intersection manifold determined by $\vec{k}=(1,1,4,6)[12]$,
corresponding to such a fibre in the mirror manifold. All these $CY_d$
spaces can be constructed by Diophantine expansions of unit monomials
$E_n=(1,...,1) \rightarrow \{P_1,P_2,P_3,P_4|1/4(P_1+P_2+P_3+P_4)=E_n\}$,
where $n = d + 2$. They can also be obtained by Diophantine expansions of
conic monomials $C_2=(2,2,...,2,0)$ appearing along the second sloping
line, together with $C_1=(0,...,0,2)$: $C_2 \rightarrow
\{P_1,P_2,P_3|1/3(P_1+P_2+P_3)=C_2\}$.  Finally, these $CY_d$ spaces can
also be obtained by Diophantine expansions of the cubic or quartic
monomials of the third $E_r$ line: $(Cub) \rightarrow
\{P_1,P_2|1/2(P_1+P_2)=(Cub)\}$. The lists of such $CY_d$ can be
determined in a similar way to the list of $CY_d$ spaces with Weierstrass
fibres, with the difference that, instead of sextic monomials, we consider
IMs of twelfth degree in any fixed variable. The corresponding formula for
the number of $CY_d$ spaces with $(1,1,4,6)[12]$ intersections is:
 
\begin{equation}
N_{\{ 39 \}_\Delta}(n) =  C_{n+3}^{n-4} = \frac{(n+3)!}{(7!)(n-4)!},
\end{equation}
where $n \geq 5$ and the index 39 corresponds to the total number of
monomials in the intersection.

\section{Conclusions and Prospects}

In previous papers~\cite{AENV1,AENV2,AENV3}, 
we have described two aspects of a Universal
Calabi-Yau Algebra that enables one in principle to classify and
enumerate Calabi-Yau spaces and their fibrations in any number of dimensions.
One aspect of this algebraic approach is the normal extension of
reflective weight vectors to higher dimensions~\cite{AENV1,AENV2}, 
and the other is the Diophantine expansion of invariant monomials,
which was first discussed in~\cite{AENV3}, and has been further developed
in this paper. As discussed here, this invariant-monomial approach enables
one to enumerate systematically the numbers of interesting fibrations of
Calabi-Yau manifolds in arbitrary dimensions.
The same approach could be adapted to discuss in a similar way
manifolds with $SO(n)$ holonomy, 
a topic extending beyond the scope of this paper.

We close by noting that the basic ideas of the Universal
Calabi-Yau Algebra can be accommodated within the
general theory of operads~\cite{Loday}. We leave this and the
relation of our work to generalizations of algebraic concepts
for a future discussion~\cite{Volkov}.

\section*{Acknowledgements}

D.V.N. acknowledges support by D.O.E. grant DE-FG03-95-ER-40917.
G.V. thanks Paul Sorba for his support while working on this paper, and 
also thanks Robert Coquereaux and Lev Lipatov for useful discussions.


\begin{thebibliography}{99}

\bibitem{CY}
E. Calabi, {\it On K\"ahler Manifolds with Vanishing Canonical Class},
in {\it Algebraic Geometry and Topology}, A Symposium in Honor
of S. Lefshetz, 1955 (Princeton University Press, Princeton, NJ, 1957); \\
S.-T. Yau, {\it Calabi's Conjecture and Some New Results in Algebraic
Geometry}, {\it Proc. Nat. Acad. Sci.} {\bf 74} (1977) 1798; \\
P. Candelas, G. Horowitz, A. Strominger and E. Witten, {\it Nucl. Phys.}
{\bf B258} (1985) 46; \\
T. H\"{u}bsch, {\it Calabi-Yau Manifolds - A Bestiary for Physicists},
(World Scientific Pub. Co., Singapore, 1992).

\bibitem{Kodaira}
M. Bershadsky, K. Intriligator, S. Kachru, D. R. Morrison, V. Sadov,
and C. Vafa, {\it Geometric Singularities and Enhanced Gauge Symmetries},
{\it Nucl. Phys.} {\bf B481} (1996) 215.

\bibitem{K3}
P. Aspinwall,  {\it K3 Surfaces and String duality}, RU-96-98,
hep-th/9611137;\\
M. Kreuzer and H. Skarke, {\it Reflexive Polyhedra, Weights and Toric
Calabi-Yau Fibrations}, HUB-EP-00/03, TUW-00/01, math.AG/0001106.\\
B. R. Greene, {\it String Theory on Calabi-Yau Manifolds}, CU-TP-812,
hep-th/9702155.

\bibitem{Skarke} M. Kreuzer and H. Skarke, {\it On the classification
of reflexive Polyhedra} {\it hep-th/9512204};
H. Skarke {\it Mod. Phys. Let.} {\bf A11} (1996) 1637,
{alg-geom/9603007}.\\
M. Kreuzer and H. Skarke,
{\it  Complete classification of reflexive polyhedra
in four dimensions }, {hep-th/0002240};
{\it Reflexive polyhedra, weights and toric Calabi-Yau
fibrations}
{\it Rev. Math. Phys.} {\bf 14} (2002) 343-374;
{\it Classification of Reflexive Polyhedra in Three Dimensions},
{hep-th/9805190}.

\bibitem{berger}
M. Berger, {\it Sur les groupes d'holonomie des vari\'et\'es \`a connexion 
affine et des vari\'et\'es riemanniennes}, Bull. Soc. Math. France {\bf 
83} (1955),279-330.

\bibitem{Joyce}
D. Joyce, {\it Compact Manifolds with Special Holonomy},
(Oxford University Press, Oxford, 2000).

\bibitem{Cartan}
E. Cartan, {\it Sur une classe remarquable d'espaces de Riemann}, 
Bull. Soc. Math. France {\bf 54} (1926), 214-264.

\bibitem{AENV1}
F. Anselmo, J. Ellis, D. V. Nanopoulos and G. G. Volkov,
{\it Towards an Algebraic Classification of Calabi-Yau Manifolds
I: Study of K3 Spaces},
{\it Phys. Part. Nucl.} {\bf 32} (2001) 318-375;
{\it Fiz. Elem. Chast. Atom. Yadra} {\bf 32} (2001) 605-698.

\bibitem{AENV2}
F. Anselmo, J. Ellis, D. V. Nanopoulos and G. G. Volkov,
{\it Results from an Algebraic Classification of Calabi-Yau
Manifolds}, {\it Phys.Lett.} {\bf B499} (2001) 187-199.

\bibitem{AENV3}
F. Anselmo, J. Ellis, D. V. Nanopoulos and G. G. Volkov,
{\it Universal Calabi-Yau algebra: Towards an unification of complex  
geometry}, CERN-TH/2001-380,
arXiv:hep-th/0207188.

\bibitem{Burris} S. Burris and H.P. Sankappnavar,
{\it A Course in Universal Algebra},  {The Millennium Edition, 2001}.

\bibitem{Batyrev}
V. Batyrev, {\it Dual Polyhedra and Mirror Symmetry for Calabi-Yau
Hypersurfaces in Toric Varieties}, {\it J. Algebraic Geom.} {\bf 3} (1994)
493; {\it Duke Math. J.} {\bf 75} (1994) 293.

\bibitem{Can1}
P. Candelas and A. Font,  {\it Duality Between the Webs of Heterotic and
Type II Vacua}, {\it Nucl. Phys.} {\bf B511} (1998) 295;\\
P. Candelas, E. Perevalov and  G. Rajesh,
{\it Toric Geometry and Enhanced Gauge Symmetry of
 F-Theory/Heterotic Vacua}, {\it Nucl. Phys.} {\bf B507} (1997) 445;\\
P. Candelas, E. Perevalov and  G. Rajesh, {\it Matter from Toric Geometry},
{\it Nucl. Phys.} {\bf B519} (1998) 225;\\
P. Candelas and H. Skarke,
{\it F-theory, SO(32) and  Toric Geometry}, {\it Phys. Lett.} {\bf B413}
(1997) 63.

\bibitem{Loday} J.-L.Loday, {La Renaissance des Operades},
{Expos\'e 792 Seminaire Bourbaki, Asterisque 1994/95}.

\bibitem{Volkov}
G. Volkov, work in preparation.

\end{thebibliography}
\end{document}